\documentstyle[psfig]{cdc5} 
 \newcounter{remark}

\begin{document}
\newcommand{\unl}{\underline}
\newcommand{\8}{\partial}
\newcommand{\dis}{\displaystyle}
\newcommand{\beq}{\begin{equation}}
\newcommand{\eeq}{\end{equation}}
\newcommand{\Cset}{{ C}\!\!\!\!{I}\;}
\newcommand{\Rset}{{\rm I}\!{\rm R}}
\newcommand{\barr}{\begin{array}}
\newcommand{\earr}{\end{array}}
\newcommand{\remark}{\vspace{1mm} \noindent \stepcounter{remark}
{\bf Remark \arabic{remark}: }}

\title{Optimal switching policies using coarse timesteppers.}
\author{
 Antonios Armaou\thanks{Department of Chemical Engineering,
The Pennsylvania State University, State College, PA 16802,
E-mail:armaou@engr.psu.edu},
 Ioannis G. Kevrekidis\thanks{Department of Chemical Engineering,
PACM and Mathematics, Princeton
University, Princeton, NJ 08544, E-mail: yannis@princeton.edu}}
\maketitle

\begin{abstract}
We present a computer-assisted approach to approximating
coarse optimal switching policies for systems described by
microscopic/stochastic evolution rules.
The ``coarse timestepper" constitutes a bridge between
the underlying kinetic Monte Carlo simulation and traditional, continuum
numerical optimization techniques formulated in discrete time.
The approach is illustrated through a simplified kinetic Monte
Carlo simulation of $NO$ reduction on a $Pt$ catalyst: a switch
between two coexisting stable steady states is implemented
by minimal manipulation of a system parameter.

\end{abstract}


\vspace*{-1mm}
\section{Introduction}\vspace*{-2mm}

The search for optimal time-varying operation protocols for
chemically reacting systems has remained an exciting research
subject for many decades.
It has been receiving increased attention recently, both for
lumped and distributed in space systems (e.g. \cite{VVB90,ac02});
as sensing and actuation become increasingly more resolved in
space and time, spatiotemporally complicated operating policies
can be considered (e.g. \cite{WPKRE01}).
Proposed computational approaches may involve (a) solution of the
temporally discretized problem (both for the process and for the
operating variable(s)) simultaneously, using large sparse linear
algebra techniques (e.g \cite{VVB90}); (b) formulations involving
direct integration of the model equations in time, keeping track
of possible constraint violations \cite{vsp94,bcvm00}; or dynamic
programming formulations.
Knowledge of a macroscopic process model, in the form
of macroscopic mass balances closed through appropriate constitutive
expressions -such as chemical kinetic rate formulas-,
is a fundamental prerequisite for these computational solution strategies.

In contemporary engineering modeling, however, we are often faced
with problems for which the available physical description is in
the form of atomistic / stochastic evolution rules (kinetic Monte
Carlo, Lattice Boltzmann, molecular dynamics, Brownian dynamics)
while the design, optimization or control is required at a
coarse-grained, macroscopic level.
Over the last few years we have been developing a computational
approach enabling microscopic / stochastic simulators to {\it
directly} perform system-level tasks, such as coarse integration,
stability analysis, bifurcation/continuation\cite{TQK00, GKT02,
KGHKRT02} and feedback control \cite{SAMK02}, thus circumventing
the derivation of explicit macroscopic evolution equations.

Here we demonstrate the extension of this computational enabling
technology to coarse optimization tasks.
In particular, we computationally approximate coarse optimal
switching policies for (the expected behavior of) a kinetic Monte
Carlo simplified model of a catalytic chemical reaction.
The system is characterized by two ``coarse-grained" stable
stationary states. We seek optimal (for a particular definition of
the cost function) parameter variation policies that will switch the kinetic
Monte Carlo simulation from one steady state to the other within a
finite time interval.

\vspace*{-1mm}
\section{Process description}\vspace*{-2mm}

We investigate microscopic/stochastic processes for which we
believe that the coarse-grained, expected dynamics can be
adequately approximated by a model of the general type
 \beq\label{process}
\dot{x} = F(x,p)
 \eeq
but where the right-hand-side of the evolution law, $F$, is
not available in closed form.
Here, $x(t) \in\Rset^n$ is a state variable vector, $t$ is
the time, $\dot{x}$ is the time derivative of $x$,
$\dis{\frac{dx}{dt}}$, and $p \in {\cal P}^m$ is the vector of
process parameters (${\cal P}^m \subset \Rset^m$ is the subset of
available values of the process parameters).
The state variables $x(t)$ are typically a few lower order moments
of an atomistically evolving distribution (e.g. a concentration,
or, in our example, a surface coverage, the zeroth moment of the
distribution of adsorbates on the surface).
The (unavailable) equation for the expected behavior of the
process may possess, at fixed process parameter values, one or
more steady states $x_{ss,i}$.

\subsection{The coarse timestepper}

The basis of our approach is the computation of a deterministic
optimal policy for the expected dynamics of the process (the
``coarse-grained dynamics") circumventing the derivation of a
closed form evolution equation for these dynamics.
This deterministic policy for the coarse-grained behavior will
then be applied to individual realizations of the process.
As we will explain below, it is convenient in our approach to reformulate
the coarse dynamics in discrete rather than continuous time.
The coarse evolution law then takes the form:
 \beq\label{proc_disc}
 \barr{l}
x_{i+1} = G_T(x_{i},p(t)),\quad t\in(t_i,t_i+T]   \vspace{2mm}\\
 \earr
 \eeq
where $x_i$ is the state at the beginning of $i$-th time interval,
$t_i$, $T$ is the time interval duration (reporting horizon),
$G_T$ represents the evolution of Eq.\ref{process} for a process
parameter profile $p(t)$, initialized at $x_i$ and evolved for
time $T$, arriving at state $x_{i+1}$ at $t_i + T \equiv t_{i+1}$.

Conventional algorithms for the solution of optimization problems
involving $G_T$ incorporate frequent calls to a subroutine that {\it evaluates}
$G_T$ and/or the action of its derivatives on initial conditions.
Equation-free based algorithms {\it estimate} the same quantities
by short bursts of (possibly ensembles of) microscopic simulations
{\it conditioned on} the same macroscopic initial conditions.
The coarse time-stepper constitutes such an {\it estimate} of the
discrete-time, macroscopic input-output map $G_T$ obtained via the kinetic
Monte Carlo simulator.
Through a {\it lifting} operator the macroscopic initial condition
is translated into several consistent microscopic initial conditions
(distributions conditioned on a few of their lower moments).
This ensemble of microscopic initial conditions is then evolved
microscopically, in an easily parallelizable fashion (one
consistent realization per CPU).
The results are averaged through a {\it restriction} operator back to
a macroscopic ``output"; it is precisely this output that traditional
algorithms simply compute through function evaluations when the
evolution equations are available in closed form.
As extensively discussed in \cite{mmk02,KGHKRT02}, part of the
microscopic evolution is spent in a ``healing" process - the
higher moments, which have been initialized ``wrong" quickly relax
to functionals of the low order moments (our state variables).
A separation of time scales (fast relaxation of the high moments
to functionals of the low ones, and slow -deterministic- evolution
of the low ones) underpins the existence of a deterministic coarse
grained evolution law. The dynamics of the evolving microscopic
distribution moments constitute thus a singularly perturbed
system.
The requirement of finite time microscopic evolution (necessary to
the moment healing process) conforms with the discrete-time
formulation of the coarse optimization problem, which is common in
many optimization algorithms (see section \ref{s:opt} below).

\subsection{Numerical experiment}\label{s:proc}

\begin{figure}[htbp]
\centerline{\psfig{file=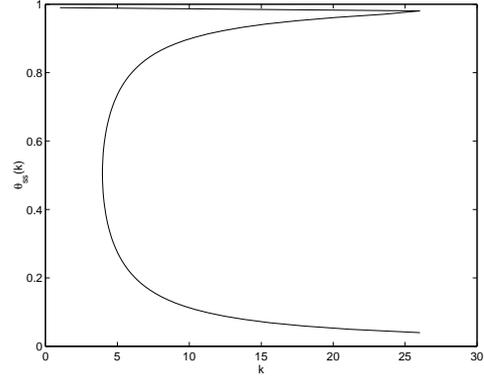,height=2.0in}}
\caption{Bifurcation diagram of $\theta$ at steady state with
respect to $k$.} \label{fig1}
\end{figure}

We illustrate the proposed combination of traditional optimization
techniques with stochastic simulators through a kinetic Monte
Carlo realization (using the stochastic simulation algorithm,
proposed by Gillespie \cite{g76,G92}) of a drastically simplified
kinetic model of $NO$ reduction by $H_2$ on Pt surfaces:
 \beq\label{rxn_ct}
 \dis{\frac{d\theta}{dt}}=\alpha(1-\theta)-\gamma
 \theta-k(1-\theta)^2 \theta.
 \eeq
Here $\theta$ describes the surface coverage of adsorbed $NO$,
$\alpha$, $\gamma$ are the $NO$ adsorption and desorption rate
constants respectively, and $k$ is the reaction rate
constant.
The reaction term is third order due to
the need for two free adjacent sites for
the adsorption of $H_2$.
In Figure
\ref{fig1} we present the deterministic bifurcation diagram
in the form of coverage $\theta_{ss}$ at
steady-state as a function of $k$ for $\alpha=1$ and
$\gamma=0.01$.
We observe a range of $k$ values for which the system exhibits
multiple steady-states; the higher and lower ones are locally
stable, while the middle one is unstable.

A description of the use of the coarse KMC timestepper in
obtaining ``coarse" versions of this bifurcation diagram
may be found in \cite{mmk02}.

\vspace*{-1mm}
\section{Coarse computational optimization}\label{s:opt}\vspace*{-2mm}

Computing the temporal profiles of the process parameters that
cause the transition of the coarse-grained system from an initial
stationary state to a different final one can be formulated as an
optimization problem; the objective is to minimize an integral
cost function over time:
 \beq\label{opt_obj_inf}
 \dis{\min_{p(t)\in {\cal P}^m}}\dis{\int_0^{\infty}}
{\cal Q}(t,x,p)
 dt
 \eeq
where ${\cal Q}$ is a continuous scalar cost function. The
constraints for this coarse optimization problem are the
(unavailable) coarse process evolution equations Eq.\ref{process},
the initialization at one of the coarse stationary states for $p =
p_{ss}$, the requirement of termination at a different stationary
state for the {\it same} process parameter value $p_{ss}$, as well
as, possibly, other inequality constraints $g(x,p)$:
 \beq\label{opt_con_inf}
 \barr{c}
 \dot{x}-F(x(t),p(t))=0, \quad g(x,p)\le 0 \vspace{2mm}\\
 \barr{ll}
 p(0)=p_{ss},& \dis{\lim_{t\rightarrow \infty}}p(t)=p_{ss} \vspace{2mm}\\
 x(0)=x_{ss,1},& \dis{\lim_{t\rightarrow \infty}}x(t)=x_{ss,2}
 \earr
 \earr
 \eeq
This is an infinite dimensional problem in continuous time. Direct
solution methods are based on the calculus of variations.
Semi-infinite programming approaches provide us with the necessary
mathematical tools to solve such problems with finite time horizon
\cite{hk93}, through discretization of the temporal domain.

Another approach consists of approximating this problem through a
finite time horizon problem with a final state penalty, which is
(in our case) subsequently solved in discrete time.
This results in a finite dimensional,
generally nonlinear, optimization problem which
-if the coarse equations are available-
could be solved using available optimization techniques.
For example, discretizing the process time $[0,t_f]$ in $N$ time
intervals of length $T$ (not necessarily constant) and assuming
that the process parameters remain constant within each interval
$p(t)=p_{i+1},\; \forall t\in(t_i,t_i+T]$, results in the
following optimization program with $(N+1)\times n+N\times m$
variables and $n\times (N+1)+m$ equality constraints:
 \beq\label{opt_fin-msm}
 \barr{c}
\barr{r@{}l} \dis{\min_{p_i\in {\cal P}^m}}& \dis{\sum_{i=1}^N}
\dis{\int_{(i-1)T}^{iT}}{\cal Q}_d(t,x_i,x_{i-1},p)dt \vspace{2mm}\\
&+{\cal W}(R(|x_N-x_{ss,2}|-\epsilon))\earr\vspace{2mm}\\
 {\rm s.t.} \vspace{2mm}\\
 g_d(x_{0},\dots,x_{N},p)\le 0,\vspace{2mm}\\
 x_{i} = G_T(x_{i-1},p)=0,\quad i=1,\dots,N,\quad  x_0=x_{ss,1} \vspace{2mm}\\
 p(t)=\dis{\sum_{i=1}^{N}p_i\Pi(\frac{t}{T}-i-\frac{1}{2})}+p_{ss}H(-t).
 \earr
 \eeq
Here ${\cal Q}_d$ is analogous to the ${\cal Q}$ function for the
discrete values of the state, $g_d$ is similarly analogous to $g$,
${\cal W}(\cdot)$ is a class $K$ scalar function, $R(\cdot)$ is
the ramp function and $\epsilon$ is a value for which
$|x_N-x_{ss,2}| \le \epsilon <=> {\cal W}=0$, $\Pi(\cdot)$ is the
standard boxcar function and $H(\cdot)$ denotes the Heaviside
function.
Appropriate final state penalty contributions to the cost function
take the place of final state constraints at infinite time as
stated in the original formulation of the problem; the final state
is restricted to be (in finite time) within a neighborhood of the
final stationary state.
This fully discrete-time formulation is ideally suited for
linking with a coarse timestepper.
The optimization problem formulated in Eq.\ref{opt_fin-msm} can
further be reduced in size by discretizing {\it only} the process
parameter temporal behavior, resulting in the following
formulation (coined {\it control vector parameterization}
\cite{vsp94})
with $N\times m$ variables and $m$ equality constraints:
 \beq\label{opt_fin-cvp}
 \barr{c}

\barr{r@{}l} \dis{\min_{p_i\in {\cal P}^m}}& \dis{\sum_{i=1}^N}
\dis{\int_{(i-1)T}^{iT}}{\cal Q}_d(t,x_i,x_{i-1},p)dt \vspace{2mm}\\
&+{\cal W}(R(|x_N-x_{ss,2}|-\epsilon))\earr\vspace{2mm}\\
 {\rm s.t.} \vspace{2mm}\\
 g_d(x_{0},\dots,x_{N},p)\le 0,\vspace{2mm}\\
p(t)=\dis{\sum_{i=1}^{N}p_i\Pi(\frac{t}{T}-i-\frac{1}{2})}+p_{ss}H(-t)
 \vspace{2mm}\\
 {\rm where} \vspace{2mm}\\
 x_{i} = G_T(x_{i-1},p)=0,\quad i=1,2,\dots,N\vspace{2mm}\\
x_0=x_{ss,1}.
 \earr
 \eeq
The difference with the previous formulation lies in that the
variables $x_{i},\;\;i=1,\dots,N$ are solved for internally,
to reduce the size of the optimization problem.
In cases where the explicit form of Eq.\ref{process} is
unavailable, the state evolution is provided through direct
simulation of the system.

\subsection{Solution methodology}\label{s:sol}

Traditional discrete time optimization schemes would call, during
the solution process, a numerical integration subroutine for the
system equations (and possibly variational integrations for the
estimation of derivatives).
This call is now substituted by the coarse timestepper; the most
important numerical issue is that of noise, inherent in the
lifting process and the stochastic simulations, and the variance
reduction necessary to estimate the state or its various derivatives.
Simulations of different physical size systems (different lattice
sizes in our simulation) are characterized by different levels of
noise, while the expectation asymptotically approaches a limiting
value for infinite system size.
For the type of simulations in this paper, changing the physical
size of the simulated domain on the one hand, and changing the number of
copies of the simulation on the other, have comparable effects in reducing the
variance of the simulation output; this, however, is not generally
the case in KMC simulations.
When a switching policy {\it for a particular physical size
system} is required (e.g. for nanoscopic reacting systems, such as
the chemical oscillations on Field Emitter tips \cite{SBJEI99})
variance reduction can be affected through a larger ensemble of
consistent microscopic initializations (see also \cite{MO95} for a
variance reduction technique using stochastic calculus).
Simulation noise affects both function evaluations and
(numerical) coarse derivative evaluations, and thus becomes
an important element of the approach.
It is interesting, however, to observe that massively parallel computation
can reduce the {\it wall clock time} required for the computation
distributing different microscopic initializations to
different CPUs.

Due to the presence of noise in the coarse timestepper results,
the use of optimization algorithms that are
specifically designed to be insensitive to noise becomes necessary
(it is well known
that the numerical estimation of derivatives is highly susceptible
to noise).
A class of direct search algorithms that fulfill this criterion
are ones that use only function evaluations to search for the
optimum such as iterative dynamic programming \cite{rflm01},
Luus-Jaakola \cite{lj73}, Nelder-Mead and Hooke-Jeeves algorithms.
Algorithms that compute local and bounded approximations of the
Jacobian matrix of the cost function with respect to the process
parameters have also been developed, such as the implicit
filtering algorithm.
The reader may refer to
\cite{kelley99} for a review of these methods.

In all the above iterative algorithms, an appropriate initial
guess of the process policy profile and a set of search directions
for the $m\times N$ variables is required. The usual set of search
directions (also used in our numerical experiment) is the unitary
basis for $\Rset^{m\times N}$.
The components of the search algorithms also
include a vector, the elements of which are the maximum distances
the algorithm should venture from the current position during the
new direction search step at each iteration. These perturbation
distances are called scales, and appear in decreasing order.

Selection of the scales in the search algorithm
requires estimates of the noise magnitude. A scale
that is too small not only leads to increased computation time
with no apparent advantages, but, especially in the case of
algorithms that compute approximations of the Jacobian, can
lead to grossly erroneous results for the search.
The following timestepper protocol yields, in our case, simulation
results of known variance magnitude:
 \vspace*{-5mm}
\begin{enumerate}
 \item Initialize timestepper. Set:
 \begin{itemize}
  \item lattice size $N_{l}$,
  \item solutions sampling size $M_{r}$,
  \item variance metric $d$ as the square root of the variance,
  subsequently
  divided by the average value of the sample.
  \item variance magnitude limit $d_{max}$ and
  \item maximum sample size $M_{max}$. 
 \end{itemize}
 \item In {\it each} time step, the timestepper:
 \begin{enumerate}
  \item simulates system for the desired parameter value $M_{r}$ times (this can
    be affected through different microscopic initializations and/or
    different random seeds in the Monte Carlo process) and
    computes $d$.
  \item If $d>d_{max}$ and $M_{r} < M_{max}$, increases $M_{r}$, repeat step (a)
  \item If $d\le d_{max}$ and $M_{r}>M_{min}$ decreases $M_{r}$,
  continue with next time step.
 \end{enumerate}
\end{enumerate}
 \vspace*{-3mm}
Adaptively adjusting the ensemble of realizations thus
helps in the choice of scales.
Optimization algorithms that are based only on function
evaluations are not guaranteed to converge to a global minimum, or
even to a minimum. This is due to the fact that we {\it do not}
compute the necessary optimality conditions in the neighborhood of
the result, and also because a search direction may have been
neglected.
Once the optimization algorithm has converged and produced a
policy profile, it is prudent to restart it with initial guess the
result of the previous search. This causes a large perturbation,
which may lead to a better optimum.
Once two consecutive runs have produced the same result, the free
variable profile is declared ``optimal over all scales"
\cite{kelley99}.

\vspace*{-1mm}
\section{Numerical Results}\label{s:results}\vspace*{-2mm}

The approach outlined in the previous section was applied towards
the computation of an optimal switching policy (between different
stationary states) in our
simplified $NO$ reduction model.
Specifically, our objective is to switch from the $\theta_{ss,s}$,
to the $\theta_{ss,f}$ locally stable stationary states
(traversing the unstable stationary state $\theta_{ss,i}$) with
the minimum possible effort.
The cost function in Eq.\ref{opt_fin-cvp} was (somewhat
arbitrarily) defined as:
 \[
\barr{l}
 {\cal Q}=(k(t)-k_{ss})^2(1-0.3e^{-t})T(\dis{\sum_{i=0}^{N}}\delta(t-iT))\vspace{2mm}\\
 {\cal W}=50[1-exp(R(|\theta(t_N)-\theta_{ss,f}|-\epsilon))]
\earr
 \]
where $\epsilon=0.05$ and $R$, $\delta$ denote the ramp and Dirac
function respectively.
We assume that the single ``manipulated variable" for this problem
(the process parameter $p$ of Eq.\ref{process}) is the reaction
rate constant $k$.
The process parameters used for the specific optimization problem
are presented in Table \ref{t:proc}.

\begin{table}[t]\vspace*{-4mm}
\caption{Process parameters}\label{t:proc} 
\begin{center}
\begin{tabular}{|c|c|cc|}
\hline
 Parameter  & Value & \multicolumn{2}{|c|}{Steady states}\\ \hline
 $T$        & 0.25 & $\theta_{ss,s}$ & 0.3301 \\
 $N$        & 20   & $\theta_{ss,i}$ & 0.6803 \\
 $t_f$      & 5    & $\theta_{ss,f}$ & 0.9896 \\
 $k_{ss}$   & 0.45 &&\\
 $\alpha$   & 1.0  &&\\
 $\gamma$   & 0.01 &&\\
\hline
 \end{tabular}
\end{center}\vspace*{-6mm}
\end{table}

KMC simulations based on the Gillespie algorithm formed
the basis of the coarse time-stepper that was used to
estimate the coarse system response.
A variety of lattice and sample sizes
were used in estimating the dynamic behavior of the system.
Hooke-Jeeves was the algorithm we chose to search for the optimal profile.
In Table \ref{t:HJ.MC-obj} we present the value of the cost
function for the computed optimal parameter profiles, through
which the effect of the error in the computed optimal profile is
implicitly quantified.
As the lattice size $N_{l}$ increases the KMC simulations expected
profiles asymptotically converge to the profile obtained at the
limit. For the simulations presented the expectation dependence to
$N_l$ was insignificant. A secondary advantage of the increased
lattice size was the variance reduction since $d \propto
N_{l}^{-1/2}$.
%
This leads, at large $N_{l}$, to results from the search for the optimal
profile of $k$ that are closer to ones obtained when we use the
timestepper of the actual deterministic problem
(for comparison purposes).
The main variance reduction parameter in the presented simulations
was the sample size $M_{r}$ used to compute the coarse response;
when $M_r$ increases in the KMC simulations, the noise magnitude
$d$ decreases, as expected, since $d \propto M_{r}^{-1/2}$.

\begin{table}[h]\vspace*{-2mm}
\caption{Hooke-Jeeves search results}\label{t:HJ.MC-obj}
\begin{center}
\begin{tabular}{l}
\begin{tabular}{|c||c|c|c|c|}
\hline
 Model  & Lattice & Runs & Objective & t$^*$ [s]\\
\hline\hline
 Legacy & $-$              & $-$   & 10.3709 & 129   \\
 KMC    & $100 \times 100$ & $200$ & 10.5418 & 674   \\
 KMC    & $200 \times 200$ & $400$ & 10.4155 & 4673  \\
 KMC    & $300 \times 300$ & $600$ & 10.3973 & 16300 \\
 KMC    & $400 \times 400$ & $800$ & 10.3815 & 38469 \\
 KMC    & $500 \times 500$ &$1000$ & 10.3811 & 75307 \\
 \hline
\end{tabular}\\
$^*$ single CPU pentium $IV$ at $2.4\; GHz$\vspace*{-6mm}
\end{tabular}\end{center}
\end{table}

\begin{figure}[htb]
\centerline{a)\psfig{file=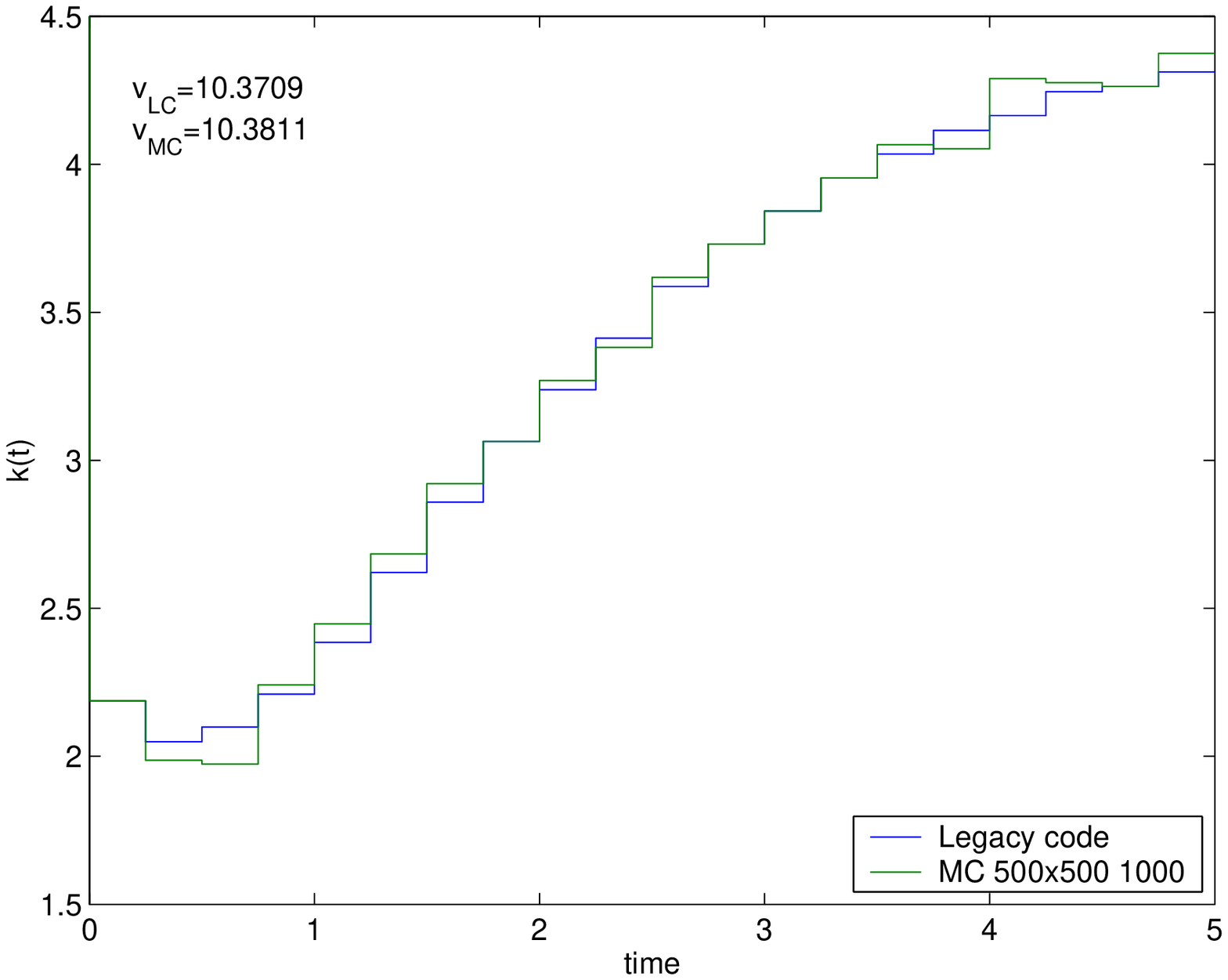,height=1.95in}}
\centerline{b)\psfig{file=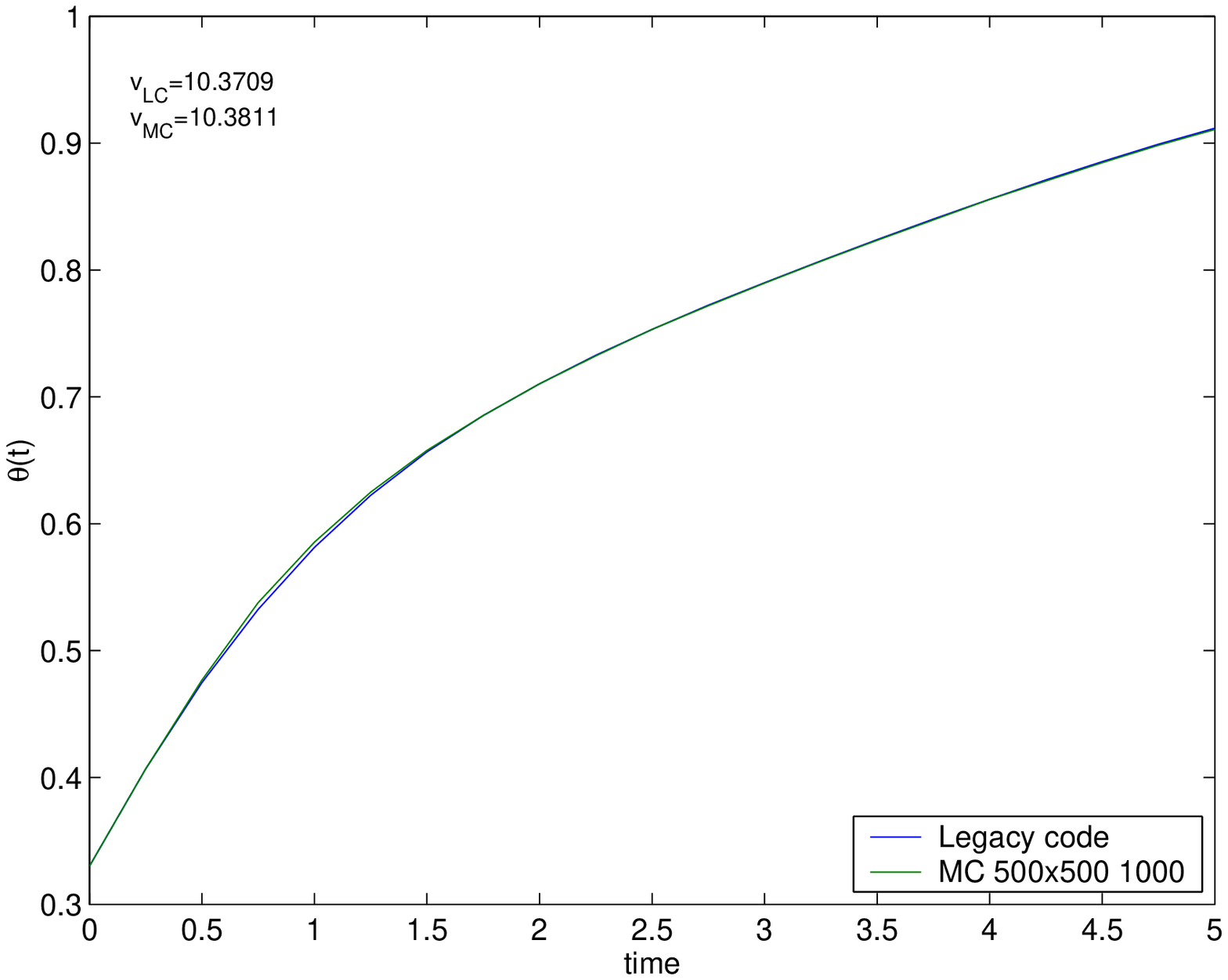,height=1.95in}}
\caption{Results using Hooke-Jeeves algorithm through numerical
integration of Eq.\ref{process} (blue line) and using KMC
simulation (green line), a) Optimal temporal profile of process
reaction rate $k$, b) Evolution of $NO$ coverage $\theta$.}
\label{figHJ2}
\end{figure}

The Gillespie algorithm was chosen so that the coarse behavior
is known at the large system size limit,
and the noisy timestepper optimization results can
be compared to it.
%
The objective value convergence to the computed
optimal value from the ODE ``direct simulator" comes at the cost of
increased computational work. The use of parallel
computing can, as we discussed, drastically decrease the necessary
wall clock time.
In Figure \ref{figHJ2} we present the results for $N_{l}=500
\times 500$ and $M_{r}=1000$ and compare them to the direct ODE
simulation results.
A {\it near}-optimal parameter profile is arrived at, due to the
combination of MC simulations' noise, system size, and the
Hooke-Jeeves search algorithm (a search direction that has been
investigated and characterized as unfavorable is not
reinvestigated to conserve CPU time).

\begin{figure*}[htb]
\centerline{a)\psfig{file=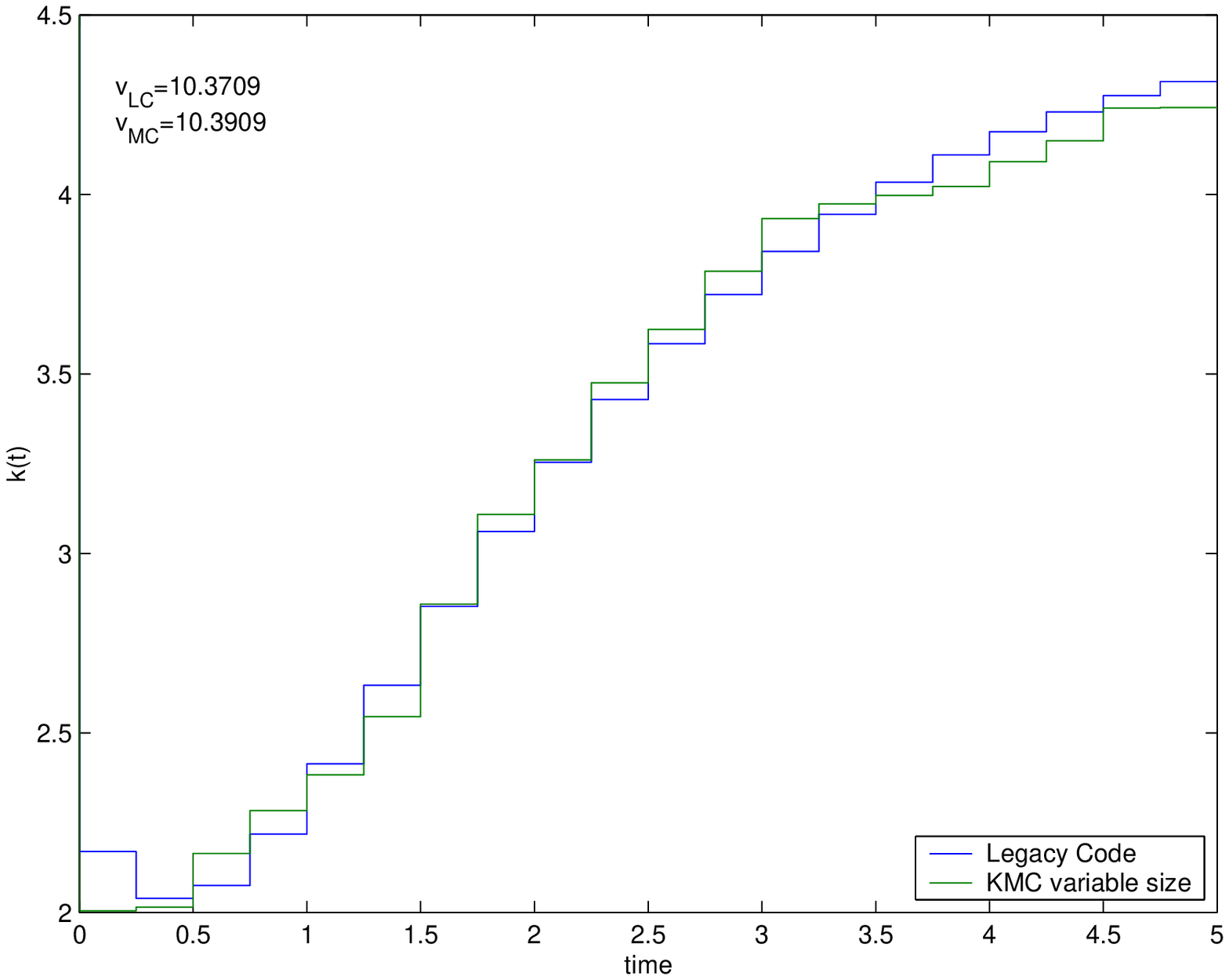,height=1.9in}
            b)\psfig{file=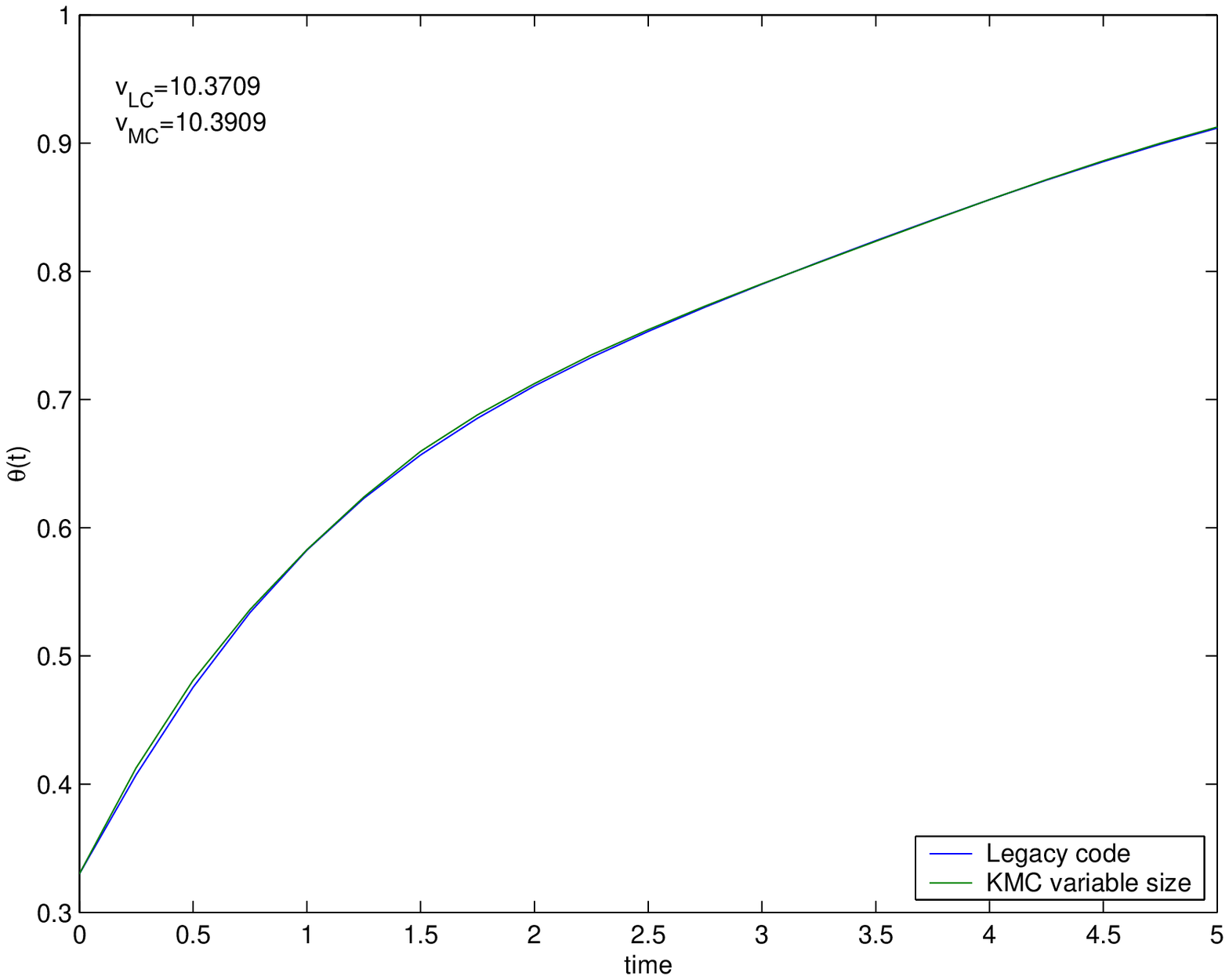,height=1.9in}
            c)\psfig{file=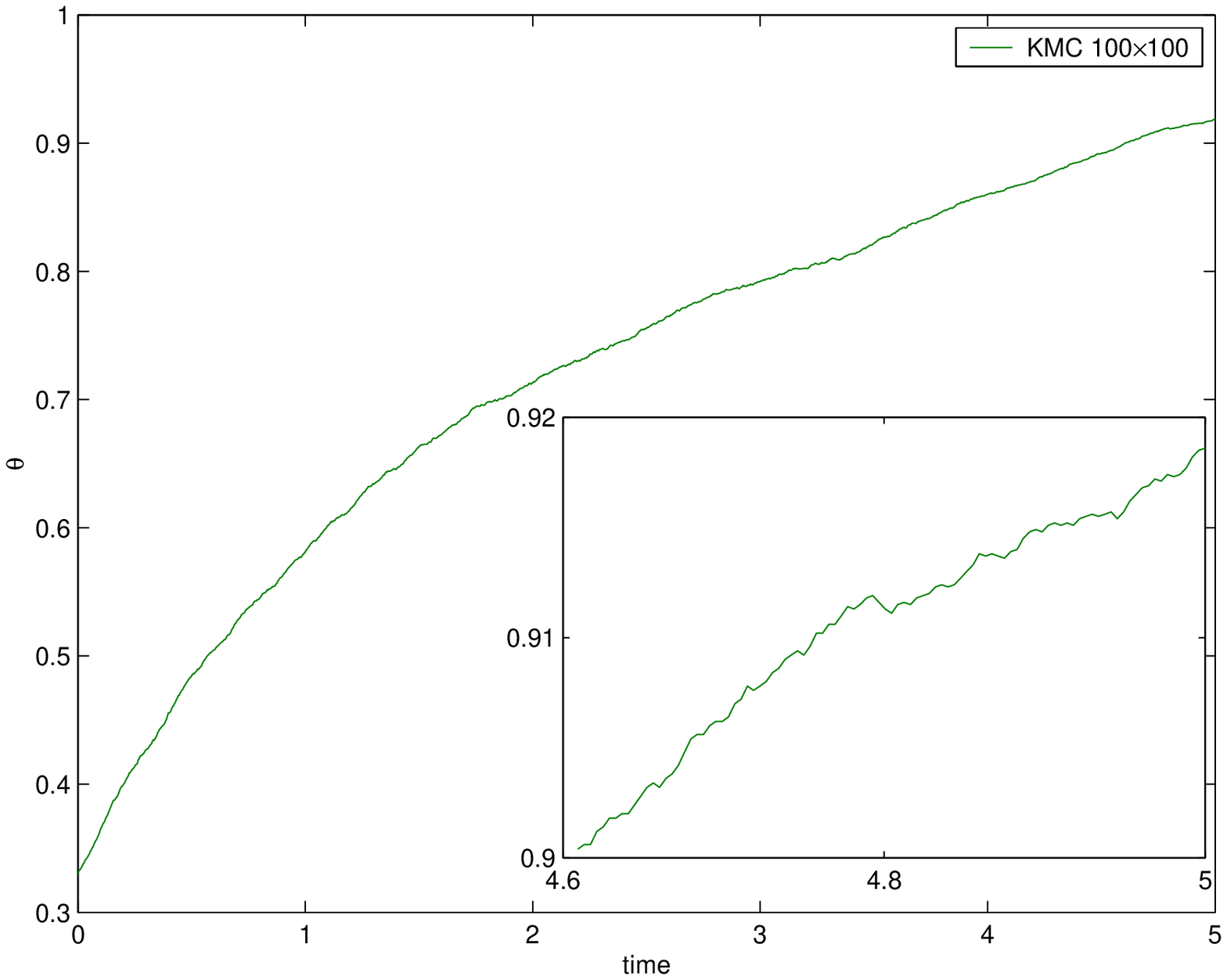,height=1.9in}}
\caption{Results using Implicit Filtering algorithm through
numerical integration of Eq.\ref{process} (blue line) and using
KMC simulation (green line), a) Optimal temporal profile of
process reaction rate $k$, b) Evolution of $NO$ coverage $\theta$,
c) Evolution of $NO$ coverage $\theta$ for a single KMC
realization, $N_l=100\times 100$, $\delta t=0.0039$.
}\vspace*{-2mm} \label{figIF1}
\end{figure*}

We also used Implicit Filtering to compute a near optimal steady
state switching profile for $k$. The method of implicit filtering
uses consecutive bounded approximations of the Jacobian with
variable step and limit to the maximum change of the design
variable at each iteration.
The timestepper protocol used had $d_{max}=0.005$,
$N_{max}=16N_{l}$, $M_{max}=4M_{r}$, $N_{l}=126\times 126$ and
$M_{r}=252$ and by adaptively adjusting $M_r$ the noise magnitude
criterion was satisfied.
The resulting near optimal time profile of $k(t)$ is presented in
Figure \ref{figIF1}a, while the near optimal path of $NO$ coverage
evolution $\theta(t)$ is shown in Figure \ref{figIF1}b for an
averaged realization, and in Figure \ref{figIF1}c for a single KMC
realization with lattice size $N_l=100\times 100$ and reporting
horizon $\delta t=0.0039$.

\begin{table}\vspace*{-2mm}
\caption{Implicit Filtering search results}\label{t:IF.MC-obj}
\begin{center}\begin{tabular}{l}
\begin{tabular}{|c||c|c|c|c|}
\hline
 Model  & $d_{max}$   & Scales                & Objective & t$^*$ [s]\\
\hline\hline
 Legacy & $-$         & $-$                   & 10.3709   & 168  \\
 KMC    & $0.005$     & $2^{-0},\dots,2^{-3}$ & 10.5461   & 297  \\
 KMC    & $0.005$     & $2^{-0},\dots,2^{-4}$ & 10.7922   & 282  \\
 KMC    & $0.0005$    & $2^{-3},\dots,2^{-5}$ & 10.4110   & 3684 \\
 KMC    & $0.001$     & $2^{-0},\dots,2^{-4}$ & 10.4129   & 6858 \\
 \hline
\end{tabular}\\
 $^*$ single CPU pentium $IV$ at $2.0\; GHz$\vspace{-8mm}
\end{tabular}\end{center}
\end{table}

In Table \ref{t:IF.MC-obj} we present computational results
obtained through incorporating the coarse timestepper in an
implicit filtering algorithm.
%
We also observe that a good scales selection depends heavily
on $d_{max}$, as scales below a certain limit have an adverse
effect on the computed near optimal path result (compare the
results of the search when the lowest scale is $2^{-4}$ to
$2^{-3}$).

Our timestepper protocol can be combined with the search
algorithm to solve successively the optimization program using
more refined scales.
Specifically, an initial search, with large values for the scales,
can take place at higher $d_{max}$, followed by searches with
gradually lower $d_{max}$ and smaller scales to refine the search
for the optimal path; this approach may lead to computational
savings.
When using the KMC with $d_{max}=0.005$ and a lower scale of
$2^{-3}$ for an initial search, followed by a KMC simulation of
the system with $d_{max}=0.0005$ and lower scale of $2^{-5}$, we
find a near optimal path with $v=10.4110$ in a total time of
$3981\;s$.
A search using KMC with
$d_{max}=0.001$ and lower scale of $2^{-4}$ lead to computing a
near optimal path of $v=10.4129$ in $6858\; s$ (the results are
shown in Table \ref{t:IF.MC-obj}).

The optimal switching path, shown in figure \ref{figIF1}b, takes
the (expected) phase point from $\theta_{ss,s}$ through the
unstable stationary state $\theta_{ss,i}$ as shown in Figure
\ref{figIF1}b.
After this is accomplished, we observe that the
optimal k(t) trajectory rapidly converges back to $k_{ss}$
(see Figure \ref{figIF1}a).
The coarse phase point is now within the region of attraction
of the steady state  $\theta_{ss,f}$, and no particular
switching action is needed to get us there.

The cost function values presented in this section were computed,
for reference purposes, by integrating the coarse system
Eq.\ref{rxn_ct} for the optimal switching profile found.
During the optimal search using KMC simulations, the coarse
process model was {\it never} used.
Several optimization algorithms were explored in conjunction with
the coarse timestepper, including Hooke-Jeeves, Nelder-Mead,
Implicit-Filtering as well as Multilevel Coordinate Search (MCS).
Hooke-Jeeves was primarily chosen due to the
simplicity of the method and its relative convergence speed.
The Implicit Filtering algorithm was mainly used with the
coarse timestepper noise-reduction protocol.
Enforcing an upper bound on the noise magnitude,
along with the selection of the value of the scales
provided an order of magnitude estimate of the error in the
estimated derivatives during the Jacobian approximation.


\vspace*{-1mm}
\section{Conclusions}\label{s:concs}\vspace*{-2mm}
We presented a computational methodology for the location of
coarse near-optimal parameter policies (in particular, steady
state switching policies) for systems for which macroscopic,
coarse evolution equations exist but are not available in closed
form.
The
advantage of the proposed method lies in the establishment, through
the coarse timestepper, of a computational bridge between atomistic/
stochastic simulators and traditional (in particular, derivative
free) optimization algorithms.
The approach can be directly extended to systems with higher
dimensional expected behavior (see for example \cite{mmk02}), and
possibly, through matrix-free methods, to systems with infinite
dimensional (spatially distributed, but dissipative) expected
behavior \cite{GKT02}.
Our current efforts focus on applying this methodology to the
study of rare events and coarse optimal paths in computational
chemistry (e.g. \cite{HK02}).

\vspace*{-3mm}
\section*{Acknowledgements}\vspace*{-2mm}
Financial support from the Air Force Office of Scientific Research
(Dynamics and Control), National Science Foundation, ITR, and the
Pennsylvania State University, Chemical Engineering Department, is
gratefully acknowledged.

\vspace*{-2mm} \begin{small}
\bibliographystyle{plain}
\bibliography{refOP}
\end{small}

\end{document}